\documentclass[onecolumn,peerreview]{IEEEtran}
\ifCLASSINFOpdf
\else
    \usepackage[dvips]{graphicx}
\fi
\usepackage{amsmath,amssymb}
\newcommand{\bysame}{%
    \leavevmode\hbox to 3em{\hrulefill}\,}

\begin{document}
%
\title{Corrections to ``An Innovations Approach to Viterbi Decoding of Convolutional Codes''}
%
%
%

\author{Masato~Tajima,~\IEEEmembership{Senior~Member,~IEEE}
\thanks{M. Tajima is with 
University of Toyama, 3190 Gofuku,
Toyama 930-8555, Japan (e-mail: masatotjm@kind.ocn.ne.jp).}
\thanks{Manuscript received April 19, 2005; revised August 26, 2015.}}

%
%

\markboth{Journal of \LaTeX\ Class Files,~Vol.~14, No.~8, August~2015}%
{Tajima \MakeLowercase{\textit{}}: Corrections to ``An Innovations Approach to Viterbi Decoding of Convolutional Codes''}
%



\maketitle

\begin{abstract}
We correct errors in Section III-B of the above-titled paper. The corrections are related to the joint distribution of the inputs to the main decoder which corresponds to a branch in the code trellis for the main decoder. Note that as far as the distribution of the input corresponding to a single code symbol is concerned, the results in Section III-B are correct. We have noticed that $\alpha$ in Proposition 12 ($\beta$ in Proposition 14) has another meaning, which has made it possible to derive the joint distribution of the inputs corresponding to the whole branch. Corrections are based on this important observation.
\end{abstract}



%
\IEEEpeerreviewmaketitle

%
%
%
%

\section{Corrections}
We correct errors in Section III-B of the above-titled paper~\cite{taji 19}.
\par
1) Eq.(62) should be replaced by
\begin{equation}
H[r_1, r_2, \cdots, r_{n_0}] \leq H[r_1]+H[r_2]+ \cdots +H[r_{n_0}] .
\end{equation}
In accordance with this correction, the paragraph prior to Eq.(62), i.e., `` Note that in our channel model, $\ldots$ the distributions corresponding to each code symbol. That is,'' should be deleted.
\par
2) In connection with Correction 1), we modify TABLE II and TABLE III. That is, we add correction terms $\Delta$ and $\tilde \Delta$, respectively. These quantities are specified in the next section.
\par
3) Eq.(71) should be replaced by
\begin{equation}
H_r^{(1)}+H_r^{(2)}+\Delta < H_{\eta}^{(1)}+H_{\eta}^{(2)}+\tilde \Delta .
\end{equation}

\section{Detailed explanations related to above corrections}
We remark that the above corrections are related to the joint distribution of the inputs to the main decoder which corresponds to a branch in the code trellis for the main decoder. As far as the distribution of the input corresponding to a single code symbol is concerned, the results in Section III-B are correct. When the paper~\cite{taji 19} was written, we thought it is difficult to obtain the joint distribution of a set of random variables $\{r_k^{(l)}\}_{l=1}^{n_0}$. Moreover, there was a misunderstanding about the independence of those random variables. After the paper was published, we have noticed that $\alpha~(=\alpha_l)$ in Proposition 12 has another meaning, which has made it possible to derive the joint distribution corresponding to the whole branch. Corrections are based on this important observation.

\subsection{Correction 1)}
Note that $\alpha~(=\alpha_l)$ in Proposition 12 is defined as
\begin{equation}
\alpha_l \stackrel{\triangle}{=}P(e_k^{(l)}=0, r_k^{(l)h}=1)+P(e_k^{(l)}=1, r_k^{(l)h}=0) .
\end{equation}
From the relation
\begin{displaymath}
\mbox{\boldmath $r$}_k^h=\mbox{\boldmath $u$}_kG+\mbox{\boldmath $e$}_k ,
\end{displaymath}
where $\mbox{\boldmath $u$}_k=\mbox{\boldmath $e$}_kG^{-1}$, and $\mbox{\boldmath $u$}_kG=\mbox{\boldmath $v$}_k=(v_k^{(1)}, \cdots, v_k^{(n_0)})$ is the encoded block for the main decoder, 
\begin{displaymath}
r_k^{(l)h}=v_k^{(l)}+e_k^{(l)}~(1 \leq l \leq n_0)
\end{displaymath}
holds. Then we have
\begin{eqnarray}
\alpha_l &=& P(e_k^{(l)}=0, r_k^{(l)h}=1)+P(e_k^{(l)}=1, r_k^{(l)h}=0) \nonumber \\
&=& P(e_k^{(l)}=0, v_k^{(l)}+e_k^{(l)}=1)+P(e_k^{(l)}=1, v_k^{(l)}+e_k^{(l)}=0) \nonumber \\
&=& P(e_k^{(l)}=0, v_k^{(l)}=1)+P(e_k^{(l)}=1, v_k^{(l)}=1) \nonumber \\
&=& P(v_k^{(l)}=1)
\end{eqnarray}
for $l=1, 2, \cdots, n_0$. Hence, the distribution of $r_k^{(l)}~(1 \leq l \leq n_0)$ is given by
\begin{eqnarray}
p_r(y) &=& (1-\alpha_l)q(y-c)+\alpha_lq(y+c) \nonumber \\
&=& P(v_k^{(l)}=0)q(y-c)+P(v_k^{(l)}=1)q(y+c) .
\end{eqnarray}
This equation means that if the code symbol is $0$, then the associated distribution obeys $q(y-c)$, whereas if the code symbol is $1$, then the associated distribution obeys $q(y+c)$. Hence, the result is quite reasonable. On the other hand, since the distribution of $r_k^{(l)}$ is given by the above equation, $r_k^{(l)}~(1 \leq l \leq n_0)$ are not mutually independent, because $v_k^{(l)}~(1 \leq l \leq n_0)$ are not mutually independent. Hence, with respect to the entropy of a set of random variables $r_k^{(1)}, \cdots, r_k^{(n_0)}$ (denoted by $H[r_1, r_2, \cdots, r_{n_0}]$), we have
\begin{displaymath}
H[r_1, r_2, \cdots, r_{n_0}] \leq H[r_1]+H[r_2]+ \cdots +H[r_{n_0}] .
\end{displaymath}
\par
We see that almost the same argument applies to $\beta~(=\beta_l)$ in Proposition 14. Note the relation
\begin{displaymath}
\mbox{\boldmath $\eta$}_{k-L}^h=u_kG+\mbox{\boldmath $e$}_{k-L} ,
\end{displaymath}
where $u_k=\mbox{\boldmath $e$}_kF$, and $u_kG=\mbox{\boldmath $v$}_k=(v_k^{(1)}, v_k^{(2)})$ is the encoded block for the main decoder. Also, note that
\begin{displaymath}
\mbox{\boldmath $\eta$}_{k-L}^h=(\zeta_k, \zeta_k) .
\end{displaymath}
Then from the relation
\begin{displaymath}
(\zeta_k, \zeta_k)=(v_k^{(1)}, v_k^{(2)})+(e_{k-L}^{(1)}, e_{k-L}^{(2)}) ,
\end{displaymath}
we have
\begin{eqnarray}
\beta_l &=& P(e_{k-L}^{(l)}=0, \zeta_k=1)+P(e_{k-L}^{(l)}=1, \zeta_k=0) \nonumber \\
&=& P(e_{k-L}^{(l)}=0, v_k^{(l)}+e_{k-L}^{(l)}=1)+P(e_{k-L}^{(l)}=1, v_k^{(l)}+e_{k-L}^{(l)}=0) \nonumber \\
&=& P(e_{k-L}^{(l)}=0, v_k^{(l)}=1)+P(e_{k-L}^{(l)}=1, v_k^{(l)}=1) \nonumber \\
&=& P(v_k^{(l)}=1)
\end{eqnarray}
for $l=1,~2$. Hence, the distribution of $\eta_{k-L}^{(l)}~(l=1, 2)$ is given by
\begin{eqnarray}
p_{\eta}(y) &=& (1-\beta_l)q(y-c)+\beta_lq(y+c) \nonumber \\
&=& P(v_k^{(l)}=0)q(y-c)+P(v_k^{(l)}=1)q(y+c) .
\end{eqnarray}
Note that (5) and (7) have the same form.

\subsection{Correction 2)}
In the following, $n_0=2$ is assumed for simplicity. Since we have already clarified the meaning of $\alpha~(=\alpha_l)$, the joint distribution of $r_k^{(1)}$ and $r_k^{(2)}$ (denoted by $p_r(x, y)$) can be derived. In fact, we have
\begin{eqnarray}
p_r(x, y) &=& \alpha_{00}q(x-c)q(y-c)+\alpha_{01}q(x-c)q(y+c) \nonumber \\
&& +\alpha_{10}q(x+c)q(y-c)+\alpha_{11}q(x+c)q(y+c) ,
\end{eqnarray}
where $\alpha_{ij}\stackrel{\triangle}{=}P(v_k^{(1)}=i, v_k^{(2)}=j)$. The associated covariance matrix is given by
\begin{eqnarray}
\Sigma_r &\stackrel{\triangle}{=}& \left(
\begin{array}{cc}
\sigma_{r_1}^2 & \sigma_{r_1r_2} \\
\sigma_{r_1r_2} & \sigma_{r_2}^2
\end{array}
\right) \nonumber \\
&=& \left(
\begin{array}{cc}
1+4c^2\alpha_1(1-\alpha_1) & 4c^2 \delta \\
4c^2 \delta & 1+4c^2\alpha_2(1-\alpha_2)
\end{array}
\right) .
\end{eqnarray}
Here, $\delta$ is defined through the following lemma.
\newtheorem{lem}{Lemma}
\begin{lem}
The following quantities have the same value:
\begin{eqnarray}
P(v_k^{(1)}=0, v_k^{(2)}=0)-P(v_k^{(1)}=0)P(v_k^{(2)}=0) &=& \alpha_{00}-(1-\alpha_1)(1-\alpha_2) \\
P(v_k^{(1)}=0)P(v_k^{(2)}=1)-P(v_k^{(1)}=0, v_k^{(2)}=1) &=& (1-\alpha_1)\alpha_2-\alpha_{01} \\
P(v_k^{(1)}=1)P(v_k^{(2)}=0)-P(v_k^{(1)}=1, v_k^{(2)}=0) &=& \alpha_1(1-\alpha_2)-\alpha_{10} \\
P(v_k^{(1)}=1, v_k^{(2)}=1)-P(v_k^{(1)}=1)P(v_k^{(2)}=1) &=& \alpha_{11}-\alpha_1 \alpha_2 .
\end{eqnarray}
The common value is denoted by $\delta$.
\end{lem}
\par
{\it Remark:} $\delta=0$ implies that $v_k^{(1)}$ and $v_k^{(2)}$ are mutually independent.
\par
\begin{IEEEproof}
From the definition of $\alpha_{ij}$, we have a system of linear equations:
\begin{eqnarray}
\alpha_{00}+\alpha_{01} &=& 1-\alpha_1 \nonumber \\
\alpha_{10}+\alpha_{11} &=& \alpha_1 \nonumber \\
\alpha_{00}+\alpha_{10} &=& 1-\alpha_2 \nonumber \\
\alpha_{01}+\alpha_{11} &=& \alpha_2 . \nonumber
\end{eqnarray}
These equations can be solved as
\begin{eqnarray}
\alpha_{00} &=& 1-\alpha_1-\alpha_2+u \nonumber \\
\alpha_{01} &=& \alpha_2-u \nonumber \\
\alpha_{10} &=& \alpha_1-u \nonumber \\
\alpha_{11} &=& u , \nonumber
\end{eqnarray}
where $u~(0 \leq u \leq 1)$ is an arbitrary constant. Hence, we have
\begin{eqnarray}
\lefteqn{\alpha_{00}-(1-\alpha_1)(1-\alpha_2)} \nonumber \\
&& =1-\alpha_1-\alpha_2+u-(1-\alpha_1-\alpha_2+\alpha_1\alpha_2) \nonumber \\
&& =u-\alpha_1\alpha_2 \nonumber \\
&& =\alpha_{11}-\alpha_1\alpha_2 . \nonumber
\end{eqnarray}
We see that the remaining three values are also equal to $u-\alpha_1\alpha_2=\alpha_{11}-\alpha_1\alpha_2$.
\end{IEEEproof}
\par
As an example, take the QLI code $C_1$ (i.e., $G(D)=(1+D+D^2, 1+D^2)$). We have
\begin{equation}
\alpha_{11}=4\epsilon-22\epsilon^2+58\epsilon^3-80\epsilon^4+56\epsilon^5-16\epsilon^6 .
\end{equation}
Since $\alpha_1$ and $\alpha_2$ have been given in~\cite{taji 19}, $\delta=\alpha_{11}-\alpha_1\alpha_2$ can be computed.
\par
Since the covariance matrix $\Sigma_r$ associated with $p_r(x, y)$ is obtained, the entropy of $(r_k^{(1)}, r_k^{(2)})$ (denoted by $H[r_1, r_2]$) is evaluated as
\begin{equation}
H[r_1, r_2]\leq \frac{1}{2}\log \bigl((2 \pi e)^2 \vert \Sigma_r \vert \bigr)
\end{equation}
with equality when $p_r(x, y)$ is Gaussian~\cite[Theorem 8.6.5]{cov 06} ($\vert \Sigma_r \vert$ denotes the determinant of $\Sigma_r$).
\par
Let $p_z(x, y)$ be the joint distribution of $z_k^{(1)}$ and $z_k^{(2)}$. We have
\begin{eqnarray}
p_z(x, y) &=& \frac{1}{4}q(x-c)q(y-c)+\frac{1}{4}q(x-c)q(y+c) \nonumber \\
&& +\frac{1}{4}q(x+c)q(y-c)+\frac{1}{4}q(x+c)q(y+c) .
\end{eqnarray}
(Note that $P(y_k^{(1)}=i, y_k^{(2)}=j)=\frac{1}{4}~(i(j)=0, 1)$.) The covariance matrix associated with $p_z(x, y)$ is given by
\begin{eqnarray}
\Sigma_z &\stackrel{\triangle}{=}& \left(
\begin{array}{cc}
\sigma_{z_1}^2 & \sigma_{z_1z_2} \\
\sigma_{z_1z_2} & \sigma_{z_2}^2
\end{array}
\right) \nonumber \\
&=& \left(
\begin{array}{cc}
1+c^2 & 0 \\
0 & 1+c^2
\end{array}
\right) .
\end{eqnarray}
Hence, the entropy of $(z_k^{(1)}, z_k^{(2)})$ (denoted by $H[z_1, z_2]$) is evaluated as
\begin{equation}
H[z_1, z_2]\leq \frac{1}{2}\log \bigl((2 \pi e)^2 \vert \Sigma_z \vert \bigr)
\end{equation}
with equality when $p_z(x, y)$ is Gaussian~\cite[Theorem 8.6.5]{cov 06}.
\par
We remark that the obtained expressions for $H[r_1, r_2]$ and $H[z_1, z_2]$ are inequalities. However, using the same argument as that in~\cite{taji 19}, we can write as follows:
\begin{eqnarray}
H[z_1, z_2]-H[r_1, r_2] &\approx& \frac{1}{2}\log \bigl((2 \pi e)^2 \vert \Sigma_z \vert \bigr)-\frac{1}{2}\log \bigl((2 \pi e)^2 \vert \Sigma_r \vert \bigr) \nonumber \\
&=& \frac{1}{2}\log \left(\frac{\vert \Sigma_z \vert}{\vert \Sigma_r \vert}\right) .
\end{eqnarray}
The above approximation is justified in the sense that the behavior of the right-hand side is consistent with the expected one of $H[z_1, z_2]-H[r_1, r_2]$, i.e., $H[z_1, z_2]-H[r_1, r_2] \geq 0$ and $H[z_1, z_2]-H[r_1, r_2]$ increases as $\epsilon$ decreases (cf. the modified TABLE II).
\par
Now the right-hand side is modified as
\begin{eqnarray}
\frac{1}{2}\log \left(\frac{\vert \Sigma_z \vert}{\vert \Sigma_r \vert}\right) &=& \frac{1}{2}\log \left(\frac{1+c^2}{1+4c^2\alpha_1(1-\alpha_1)}\right) \nonumber \\
&& +\frac{1}{2}\log \left(\frac{1+c^2}{1+4c^2\alpha_2(1-\alpha_2)}\right)+\Delta ,
\end{eqnarray}
where
\begin{equation}
\Delta\stackrel{\triangle}{=}\frac{1}{2}\log \left(\frac{\sigma_{r_1}^2 \sigma_{r_2}^2}{\sigma_{r_1}^2 \sigma_{r_2}^2-(\sigma_{r_1r_2})^2}\right) .
\end{equation}
In~\cite{taji 19}, the notations
\begin{displaymath}
H_r^{(1)}\stackrel{\triangle}{=}H[z_1]-H[r_1]\approx \frac{1}{2}\log \left(\frac{1+c^2}{1+4c^2 \alpha_1(1-\alpha_1)}\right)
\end{displaymath}
\begin{displaymath}
H_r^{(2)}\stackrel{\triangle}{=}H[z_2]-H[r_2]\approx \frac{1}{2}\log \left(\frac{1+c^2}{1+4c^2 \alpha_2(1-\alpha_2)}\right)
\end{displaymath}
are used. In order to maintain consistency as much as possible, we will continue to use the above notations after corrections. Then we have
\begin{equation}
H[z_1, z_2]-H[r_1, r_2] \approx H_r^{(1)}+H_r^{(2)}+\Delta .
\end{equation}
Observe that the derived entropy is modified by the value of $\Delta$.
\par
The above argument equally applies to $\mbox{\boldmath $\eta$}_{k-L}=(\eta_{k-L}^{(1)}, \eta_{k-L}^{(2)})$. In this case, we have
\begin{equation}
H[z_1, z_2]-H[\eta_1, \eta_2] \approx H_{\eta}^{(1)}+H_{\eta}^{(2)}+\tilde \Delta .
\end{equation}
The correction term $\tilde \Delta$ is defined by
\begin{equation}
\tilde \Delta\stackrel{\triangle}{=}\frac{1}{2}\log \left(\frac{\sigma_{\eta_1}^2 \sigma_{\eta_2}^2}{\sigma_{\eta_1}^2 \sigma_{\eta_2}^2-(\sigma_{\eta_1\eta_2})^2}\right) ,
\end{equation}
where
\begin{eqnarray}
\sigma_{\eta_1}^2 &=& 1+4c^2\beta_1(1-\beta_1) \\
\sigma_{\eta_2}^2 &=& 1+4c^2\beta_2(1-\beta_2) \\
\sigma_{\eta_1\eta_2} &=& 4c^2 \tilde \delta \\
\tilde \delta &=& \beta_{11}-\beta_1\beta_2~~(\beta_{11}\stackrel{\triangle}{=}P(v_k^{(1)}=1, v_k^{(2)}=1)) .
\end{eqnarray}
\par
As an example, consider the QLI code $C_1$ again. We have
\begin{equation}
\beta_{11}=4\epsilon-20\epsilon^2+48\epsilon^3-64\epsilon^4+48\epsilon^5-16\epsilon^6 .
\end{equation}
Since $\beta_1$ and $\beta_2$ have been given in~\cite{taji 19}, $\tilde \delta=\beta_{11}-\beta_1\beta_2$ can be computed.
\par
Based on the above results, we have modified TABLE II and TABLE III. The modified tables are given below, where the variables $\alpha_1$, $\alpha_2$, $\beta_1$, and $\beta_2$ are omitted for simplicity.
\addtocounter{table}{1}
\begin{table}[tb]
\caption{Entropies associated with input distributions (as a general code)}
\label{Table 2}
\begin{center}
\begin{tabular}{c*{6}{|c}}
$E_b/N_0~(\mbox{dB})$ & $c$ & $\epsilon$ & $H_r^{(1)}$ & $H_r^{(2)}$ & $\Delta$ & $H_r^{(1)}+H_r^{(2)}+\Delta$ \\
\hline
$0$ & $1.000$ & $0.1587$ & $0.0055$ & $0.0026$ & $0.0113$ & $0.0194$ \\
$1$ & $1.122$ & $0.1309$ & $0.0136$ & $0.0075$ & $0.0223$ & $0.0434$ \\
$2$ & $1.259$ & $0.1040$ & $0.0307$ & $0.0190$ & $0.0380$ & $0.0877$ \\
$3$ & $1.413$ & $0.0788$ & $0.0639$ & $0.0445$ & $0.0582$ & $0.1666$ \\
$4$ & $1.585$ & $0.0565$ & $0.1214$ & $0.0929$ & $0.0786$ & $0.2929$ \\
$5$ & $1.778$ & $0.0377$ & $0.2131$ & $0.1759$ & $0.0935$ & $0.4825$ \\
$6$ & $1.995$ & $0.0230$ & $0.3456$ & $0.3027$ & $0.0952$ & $0.7435$ \\
$7$ & $2.239$ & $0.0126$ & $0.5191$ & $0.4756$ & $0.0802$ & $1.0749$ \\
$8$ & $2.512$ & $0.00600$ & $0.7241$ & $0.6870$ & $0.0520$ & $1.4631$ \\
$9$ & $2.818$ & $0.00242$ & $0.9355$ & $0.9103$ & $0.0229$ & $1.8687$ \\
$10$ & $3.162$ & $0.00078$ & $1.1266$ & $1.1131$ & $0.0056$ & $2.2453$
\end{tabular}
\end{center}
\end{table}
\begin{table}[tb]
\caption{Entropies associated with input distributions (as a QLI code)}
\label{Table 3}
\begin{center}
\begin{tabular}{c*{6}{|c}}
$E_b/N_0~(\mbox{dB})$ & $c$ & $\epsilon$ & $H_{\eta}^{(1)}$ & $H_{\eta}^{(2)}$ & $\tilde \Delta$ & $H_{\eta}^{(1)}+H_{\eta}^{(2)}+\tilde \Delta$ \\
\hline
$0$ & $1.000$ & $0.1587$ & $0.0026$ & $0.0119$ & $0.0260$ & $0.0405$ \\
$1$ & $1.122$ & $0.1309$ & $0.0075$ & $0.0252$ & $0.0427$ & $0.0754$ \\
$2$ & $1.259$ & $0.1040$ & $0.0190$ & $0.0498$ & $0.0647$ & $0.1335$ \\
$3$ & $1.413$ & $0.0788$ & $0.0445$ & $0.0926$ & $0.0896$ & $0.2267$ \\
$4$ & $1.585$ & $0.0565$ & $0.0929$ & $0.1602$ & $0.1115$ & $0.3646$ \\
$5$ & $1.778$ & $0.0377$ & $0.1759$ & $0.2602$ & $0.1234$ & $0.5595$ \\
$6$ & $1.995$ & $0.0230$ & $0.3027$ & $0.3975$ & $0.1187$ & $0.8189$ \\
$7$ & $2.239$ & $0.0126$ & $0.4756$ & $0.5694$ & $0.0951$ & $1.1401$ \\
$8$ & $2.512$ & $0.00600$ & $0.6870$ & $0.7654$ & $0.0584$ & $1.5108$ \\
$9$ & $2.818$ & $0.00242$ & $0.9103$ & $0.9634$ & $0.0248$ & $1.8985$ \\
$10$ & $3.162$ & $0.00078$ & $1.1131$ & $1.1406$ & $0.0058$ & $2.2595$
\end{tabular}
\end{center}
\end{table}

\subsection{Correction 3)}
In accordance with Correction 2), Eq.(71) should be replaced by
\begin{displaymath}
H_r^{(1)}+H_r^{(2)}+\Delta < H_{\eta}^{(1)}+H_{\eta}^{(2)}+\tilde \Delta .
\end{displaymath}

\ifCLASSOPTIONcaptionsoff
  \newpage
\fi


\begin{thebibliography}{99}
\bibitem{taji 19}M.~Tajima, ``An innovations approach to Viterbi decoding of convolutional codes,'' {\em IEEE Trans.~Inform.~Theory}, vol.~65, no.~5, pp.~2704--2722, May 2019.
\bibitem{cov 06}T.~M.~Cover and J.~A.~Thomas, {\em Elements of Information Theory}, 2nd ed. Hoboken, NJ, USA: John Wiley \& Sons, 2006.

\end{thebibliography}
\end{document}